# Quantum phase interference in a fullerene-based molecular qutrit

Ye-Xin Wang[a, †], Zheng Liu[a, †], Yu-Hui Fang[a], Shen Zhou[b, d, *], Shang-Da Jiang[a, b, *], Song Gao[a, b, c, *]

[a] Beijing National Laboratory of Molecular Science, Beijing Key Laboratory of Magnetoelectric Materials and Devices, College of Chemistry and Molecular Engineering, Peking University, Beijing 100871, China
[b] School of Chemistry and Chemical Engineering, South China University of Technology, Guangzhou 510640, China
[c] Beijing Academy of Quantum Information Sciences, West Bld.#3, No.10 Xibeiwang East Rd. Beijing 100193, China
[d] College of Aerospace Science and Engineering, National University of Defense Technology Changsha 410000, P. R. China

E-mail: zhoushen@nudt.edu.cn, jiangsd@pku.edu.cn, gaosong@pku.edu.cn
† These authors contributed equally.

Supporting information for this article is given via a link at the end of the document.

**Abstract:** High spin magnetic molecules are promising candidates for quantum information processing because they intrinsically have multiple sublevels for information storage and computational operations. However, due to their susceptibility to the environment and limitation from the selection rule, the arbitrary control of the quantum state of a multilevel system on a molecular and electron spin basis has not been realized. Here we exploit the photoexcited triplet of $C_{70}$ as a molecular electron spin qutrit. After the system was initialized by photoexcitation, we prepared it into representative three-level superposition states characteristic of the qutrit, measured their density matrices, and showed the interference of the quantum phases in the superposition. The interference pattern is further interpreted as a map of evolution through time under different conditions.

## Introduction

Electron spins in molecular systems has been regarded as a promising platform for quantum information processing (QIP), because the scalability issue challenging other proposed quantum systems can be potentially dealt with by the vastly available chemistry-based bottom-up tools to manipulate molecules.[1-3] While an increasing number of new candidate molecules showing sufficiently long spin relaxation times have been synthesized recently,[4] fullerenes are still advantageous because their spin coherence is inherently preserved, which is associated with the clean intramolecular magnetic environment of carbon nuclei and the near-spherical symmetry of the cage configuration.[5-8] As a result, regarding fullerene-based QIP, ultrafast control of nuclear spin qubits assisted by electron spin,[9,10] entanglement between electron and nuclear spins,[11,12] electron spin dipolar coupling[13,14] and fullerene-based atomic clock transition,[15] has been studied experimentally.

One of the most promising advantages of magnetic molecules is their ease of being scaled up and made into spin systems of higher nuclearity or dimension, which can be used to encode more information.[16] However, while most of recent reports concerning QIP based on magnetic molecules presented different ways to use the spin as a qubit, the multilevel nature of $S > 1/2$ systems has not been thoroughly exploited yet. It has been proposed that the quantum searching algorithm can be implemented in high spin molecular magnets.[17] Since then, a variety of high spin systems on which researchers demonstrated coherent manipulations addressing the allowed transitions has been reported.[18-20] Nevertheless, a full control of a high spin system requires superposing the spin states between non-adjacent levels. Herein, utilizing photoexcited triplet state of $C_{70}$ ($S = 1$), we experimentally demonstrated the full control of the electron spin system in the excited state of fullerene, by preparing a superposition state regarding the forbidden transition and a superposition state involving the full Hilbert space basis set. With the properly tunable quantum spin system having three eigenstates (qutrit), we illustrated the interference between quantum phases of different eigenstates. Combining the interference with the concept of time proportional phase increment (TPPI),[11] we further demonstrated that a qutrit spin system can undergo an incommensurate evolution. To the best of our knowledge, this work is the first to report the full control of the electron spin in a $S > 1/2$ molecular system. $C_{70}$ is chosen for the experiments due to the following reasons. (I) It is spin-free in ground state, and an excited triplet state with a long lifetime can be reached by laser excitation and intersystem crossing.[21] This $S = 1$ system has a higher dimensional Hilbert space than the molecules proposed as qubits. (II) In the family of fullerenes, the zero-field splitting (ZFS) of this photoexcited triplet varies with the size and structure of the molecule. The ZFS of $C_{70}$ has desired magnitude, which allows us to address each transition specifically, and still falls within the bandwidth of the spectrometer. (III) In this all-carbon molecule the electron spin can have an environment relatively clean of nuclear spins, allowing slow longitudinal ($T_1$) and transverse ($T_2$) relaxation. Specifically for this photoexcited state of $C_{70}$, $T_1 = 10.7 \pm 0.2$ ms and $T_2 = 9.4 \pm 0.1$ μs (figures S1 and S2). That $T_1 \gg T_2$ enabled us to prepare and manipulate superposition states with specific quantum phases before the system significantly relaxes towards thermal equilibrium.

## Results

**Energy levels.** The energy levels of the photoexcited triplet state of $C_{70}$ involved in the experiments are illustrated in figure 1. The splittings are generated by the Hamiltonian

$$H = \mu_B \vec{B}_0^T \cdot \bar{\bar{g}} \cdot \hat{\vec{S}} + \hat{\vec{S}}^T \cdot \bar{\bar{D}} \cdot \hat{\vec{S}} \qquad (1)$$

where $\mu_B$ is the Bohr magneton, $\bar{\bar{g}}$ is the g-tensor, $\vec{B}_0$ is the applied magnetic field, $\hat{\vec{S}}$ is the spin operator and $\bar{\bar{D}}$ is the ZFS tensor. By simulation of the echo detected field sweep (EDFS) pulsed electron paramagnetic resonance (EPR) spectrum in figure 1 (inset), an isotropic $\bar{\bar{g}}$ with $g = 2.0037$ and an almost planar $\bar{\bar{D}}$ with $|D| = 152$ MHz and $|E| = 50.4$ MHz, which is in line with reported values,[22] are derived. The spectrum was taken at $f_0 = 9.2505$ GHz.

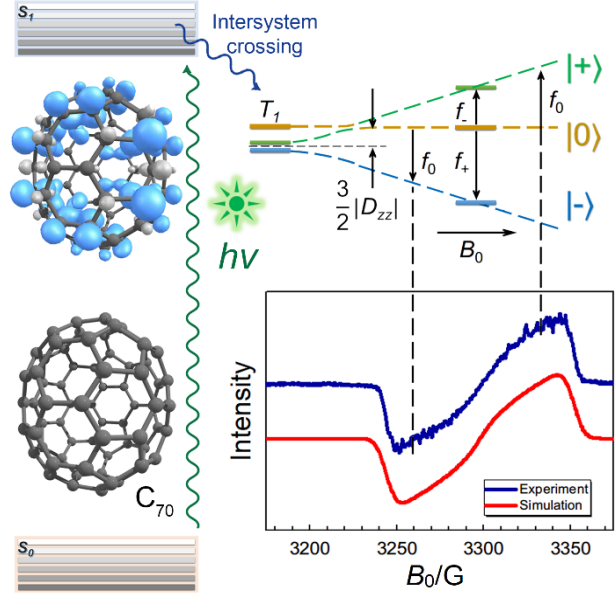

**Figure 1.** Energy levels involved in the experiments with the molecular structure and calculated spin density of the photoexcited triplet of $C_{70}$. The inset is the measured and simulated EDFS spectrum. The vertical dashed lines mark the selected portion of molecules.

The $C_{70}$ solution sample was cooled to 5 K and irradiated by a beam of 532 nm laser. The molecules were excited from its singlet ground state ($S_0$) to a singlet excited state ($S_1$) and then a triplet state ($T_1$) is reached through intersystem crossing.[21] With high-field approximation we denote the sublevel with $M_S$ = +1, 0 and -1 by |+⟩, |0⟩ and |-⟩ respectively. The spin density of this triplet state is calculated and shown in Figure 1.

In the EDFS spectrum, the downward and upward peaks correspond to resonant emission (|0⟩ → |-⟩) and absorption (|0⟩ → |+⟩) respectively. The centrosymmetric shape indicated equal population differences regarding transitions |0⟩ → |±⟩. Therefore the photoexcitation initialized the system to a pseudopure state of |0⟩, with equal populations of |±⟩.

Regarding different molecules in the ensemble, the principal axis of $\bar{\bar{D}}$ tensor was randomly oriented with respect to the z-axis of the lab frame, which was the direction of $\vec{B}_0$. In the procedure of quantum state preparation and manipulation, however, it is important that a well-defined portion in all the molecules is addressed by all the microwave pulses. We selected those with their principal axes of $\bar{\bar{D}}$ tilting about 40° from $\vec{B}_0$, corresponding to a $D_{zz}$ component of 60 MHz. This portion of molecules signaled at $B_0 = 3299 \pm 64$ G, as marked by the vertical dashed lines in figure 1. The experiments were carried out at $B_0 = 3299$ G, where these molecules can be addressed by microwave frequencies $f_\pm = f_0 \mp 90$ MHz (corresponding to transitions |0⟩ → |±⟩ respectively). These frequencies prove suitable for our experiment because they were neither too close so that the signals would be too weak, nor too far apart so that they would fall out of the bandwidth. The signals of these transitions were both detected off-resonantly at $f_0$, simplifying the experimental procedure.

**Superposition state preparation and tomography.** To demonstrate the capability to reach an arbitrary superposition state and thus qualify the system as a qutrit, we prepared the system into $|\psi_1\rangle = (|+\rangle + |-\rangle)/\sqrt{2}$, with its components only directly addressable by a forbidden transition, and $|\psi_2\rangle = (|+\rangle + |0\rangle + |-\rangle)/\sqrt{3}$, with its components covering the full Hilbert space basis set. In this context, the experimental setting described above is advantageous in that |+⟩ and |-⟩ can be treated equally, simplifying the operation (see SI) and minimizing error accumulation.

Figure 2(a) illustrates the preparation and density matrix tomography procedure of $|\psi_1\rangle$ and $|\psi_2\rangle$. After laser initialization, the system was manipulated with the preparation pulses. $|\psi_1\rangle$ was prepared with a π/2 pulse at $f_+$ followed by a π pulse at $f_-$, and $|\psi_2\rangle$ with an $\arccos(1/3)$ pulse at $f_+$ followed by a π/2 pulse at $f_-$.

After that, the real and imaginary parts of each density matrix element was determined by applying a unitary transformation to move it to the diagonal and then measuring the population differences between |+⟩ and |0⟩, or |0⟩ and |-⟩. The measurement was done by a sequence consisting of a delay for 40 μs ($> 4T_2$) to let the system decohere, a soft π/2 pulse at $f_+$ or $f_-$ and a free induction decay (FID) detection. Operational and theoretical details about the procedure are presented in SI.

As described above, the density matrices of $|\psi_1\rangle$ and $|\psi_2\rangle$ are determined to be

$$\rho_1 = \begin{pmatrix} 0.50 & -0.05 + 0.02i & 0.29 \\ -0.05 - 0.02i & 0.04 & -0.02 + 0.04i \\ 0.29 & -0.02 - 0.04i & 0.45 \end{pmatrix} \quad (2)$$

$$\rho_2 = \begin{pmatrix} 0.37 & 0.27 + 0.03i & 0.28 + 0.04i \\ 0.27 - 0.03i & 0.31 & 0.29 + 0.01i \\ 0.28 - 0.04i & 0.29 - 0.01i & 0.33 \end{pmatrix} \quad (3)$$

as shown in Figure 2(b-e).

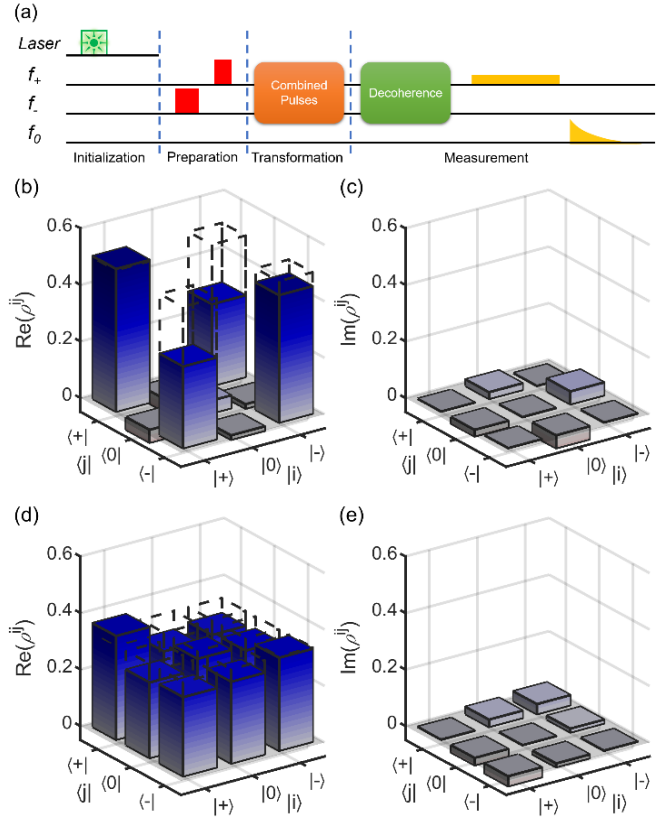

**Figure 2.** Tomography of the density matrix. (a) Pulse sequence for the preparation and density matrix tomography of $|\psi_1\rangle$ and $|\psi_2\rangle$. (b-e) Real and imaginary parts of the density matrices of (b, c) $|\psi_1\rangle$ and (d, e) $|\psi_2\rangle$, with ideal values shown by dashed bars.

With $\mathrm{Tr}(\rho_{1,2}^2) < 1$, both prepared states were apparently mixed. The fidelities of the two states were $F(\rho_1, \sigma_1) = 0.766$ and $F(\rho_2, \sigma_2) = 0.890$, where

$$F(\rho_k, \sigma_k) = \left[\mathrm{Tr}\left(\sqrt{\sqrt{\sigma_k}\rho_k\sqrt{\sigma_k}}\right)\right]^2 \tag{4}$$

Herein $k = 1, 2$, and $\sigma_{1,2}$ are the ideal density matrices. While previous reports of manipulating the triplet state of electron spin in a molecular system are rare, this fidelity is considerably better than that of the nuclear spin state prepared using the photoexcited triplet of a fullerene derivative.[10] The deviation from ideal values resulted from pulse imperfection and microwave field inhomogeneity.

**Quantum phase interference.** Figure 3 depicts quantum phase interference experiments on $|\psi_1\rangle$ and $|\psi_2\rangle$, resulting from their 3-level nature. The pulse sequence was as illustrated in figure 3(a). The preparation pulses at $f_\pm$ were shifted in phase by $\phi_\pm$ respectively, generating superposition states with the same amplitudes and different phase factors. Then the reversion pulses, which have opposite tipping angles with respect to the unshifted preparation pulses were applied in reversed order. Afterwards, the magnetization regarding $|0\rangle$ and $|+\rangle$, $M^{0+}$, was measured. It is observed that $\phi_+$ and $\phi_-$ interfere to make the magnetization exhibit a pattern characteristic of the superposition. A more detailed description is available in SI. Figure 3(b, c) show these patterns of the prepared states (left) and simulated results (right) for $|\psi_1\rangle$ and $|\psi_2\rangle$. Ideally, the magnetization would be

$$M_1^{0+}(\phi_+, \phi_-) = -\frac{1}{2}\cos(\phi_+ + \phi_-) \tag{5}$$

$$M_2^{0+}(\phi_+, \phi_-) = -\frac{4}{9}\cos\phi_+ - \frac{1}{9}\cos\phi_- - \frac{4}{9}\cos(\phi_+ + \phi_-) \tag{6}$$

To elucidate the periodicity of the patterns, their 2D fast Fourier transforms (2D-FFT) are shown in figure 3(d, e) with the magnitudes normalized. The solid line bars represent data from experiments and dashed line bars give calculated results. They are labeled with the frequencies $k_\pm$ they represent, and symmetric about point (0,0) by principle. Those with $(2\pi k_+, 2\pi k_-) = \pm(1,1)$ (A1 and A2) are a sign for the superposition of $|-\rangle$ and $|+\rangle$ states, which corresponds to a spin-forbidden transition, not present in two-level systems and characteristic of this $S = 1$ qutrit. Bars with $(2\pi k_+, 2\pi k_-) = (\pm 1, 0)$ and $(0, \pm 1)$ (B and C) marks the superposition of $|0\rangle$ with $|+\rangle$, and $|0\rangle$ with $|-\rangle$. The perfect $|\psi_1\rangle = (|+\rangle + |-\rangle)/\sqrt{2}$ will have A1 only, and $|\psi_2\rangle = (|+\rangle + |0\rangle + |-\rangle)/\sqrt{3}$ will have A2, B and C with relative heights 4:4:1. These were closely reproduced by the experiment. The imperfections resulted from the inhomogeneity of microwave field, errors in pulse lengths, and the decoherence due to non-zero operation times and pulse bandwidths.

The interference pattern can be viewed from another perspective in analogy with TPPI,[11] *i.e.* by transforming the phase coordinates into those of time. In a rotating frame with frequency $f_0$, the phase evolution of the system would be along a linear path in the $(\phi_+, \phi_-)$ plane with slope $-\Delta f_+/\Delta f_-$ where $\Delta f_\pm = \pm(f_0 - f_\pm)$. Going along this line, $M^{0+}(\phi_+, \phi_-)$ gives a function of time characterizing the evolution of the prepared superposition state.

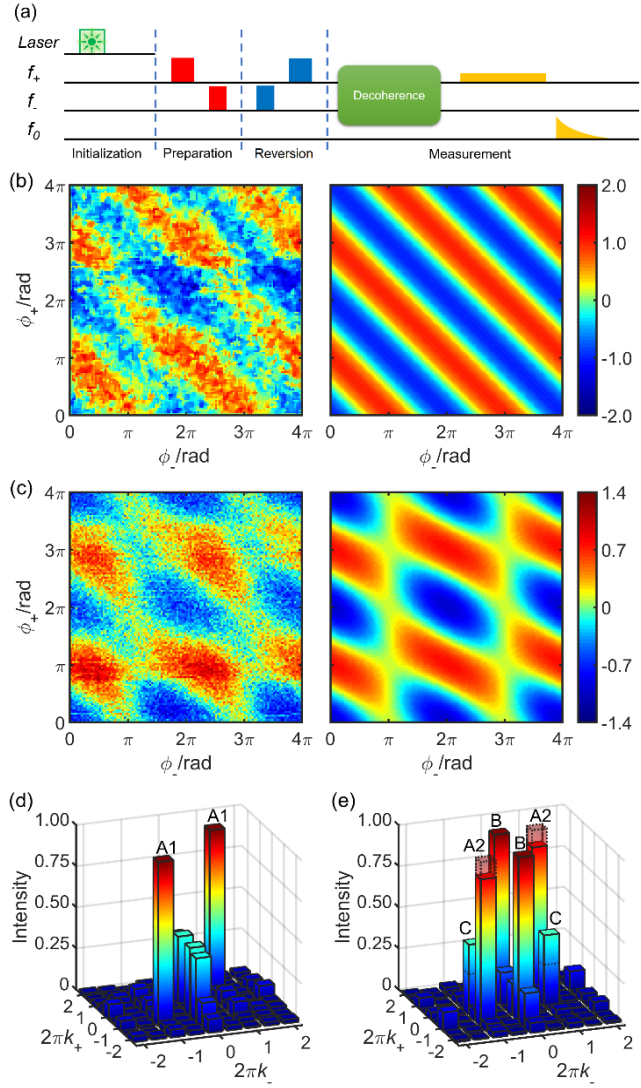

**Figure 3.** quantum phase interference experiments on $|\psi_1\rangle$ and $|\psi_2\rangle$. (a) The pulse sequence. (b, c) Measured (left) and simulated (right) patterns of $M^{0+}$ as a function of $\phi_+$ and $\phi_-$ for $|\psi_1\rangle$ and $|\psi_2\rangle$. (d, e) The 2D Fourier transformations of interference patterns (b) and (c), with solid bars showing the experimental data and dashed-line ones the simulation results. For clarity, only low-frequency data are shown and the heights are normalized.

Therefore, the coordinates of this 2D interference pattern can be changed into times of fictitious evolutions

$$t^{0\pm} = \pm\phi_\pm/\Delta f_\pm \tag{7}$$

Taking figure 3(c) as an example, having its coordinates converted accordingly we get the upper panel of figure 4(a). Its diagonal, with $t^{0+} = t^{0-} = t$, marks the actual path of evolution through time $t$. The lower panel shows how the magnetization would change.

In the aforementioned experiments, the condition is selected so that $\Delta f_+ = \Delta f_-$. However, by adjusting $B_0$ and $f_\pm$ according to figure 1, $\Delta f_+/\Delta f_-$ can be of any value. As long as $\Delta f_+ + \Delta f_- = -3D_{zz}$ still holds, the portion of molecules addressed would be the same. The interference pattern, rescaled in different ways, could cover all these cases as exemplified in figure 4(b) with $\Delta f_+/\Delta f_- = 0.5$ and 4(c) with $\Delta f_+/\Delta f_- = 2$. Generally, the evolution path of the system with any $\Delta f_+/\Delta f_-$ is represented by a horizontal line in figure 4(d). Therefore within our experimental setup, a state $|\psi\rangle = c_+|+\rangle + c_0|0\rangle + c_-|-\rangle$ can be made into having any combination of phase factors, like $|\psi'\rangle = c_+\mathrm{Exp}(i\phi_1)|+\rangle + c_0|0\rangle + c_-\mathrm{Exp}(i\phi_2)|-\rangle$, by setting $\Delta f_{+,-} = -3D_{zz}\phi_{1,2}/(\phi_1 + \phi_2)$, and waiting for $t = -(\phi_1 + \phi_2)/3D_{zz}$.

It shall be noted that the collection of all these states, parameterized by the phases $\phi_1$ and $\phi_2$, can be represented by a torus, as shown in figure 4(e). In this representation, $\phi_1$ spans along the horizontal circles going around the torus and $\phi_2$ spans along the ones perpendicular to them. The path of evolution can then be viewed as a curve twining on the torus. In ideal cases where $\Delta f_+/\Delta f_-$ are rational numbers, the paths are closed, as in the upper part (blue, red and green curves for $\Delta f_+/\Delta f_- = 0.5$, 1 and 2 respectively). When $\Delta f_+/\Delta f_-$ is irrational, the paths would never return to where it started, as depicted in the lower part. This means that a 3-state quantum system, unlike those with only 2 states, would undergo non-periodic evolution, and the magnetization would fluctuate along an incommensurate curve, although approximate periodicity in a finite amount of time can also be achieved if the off-resonant frequencies are set with enough precision.

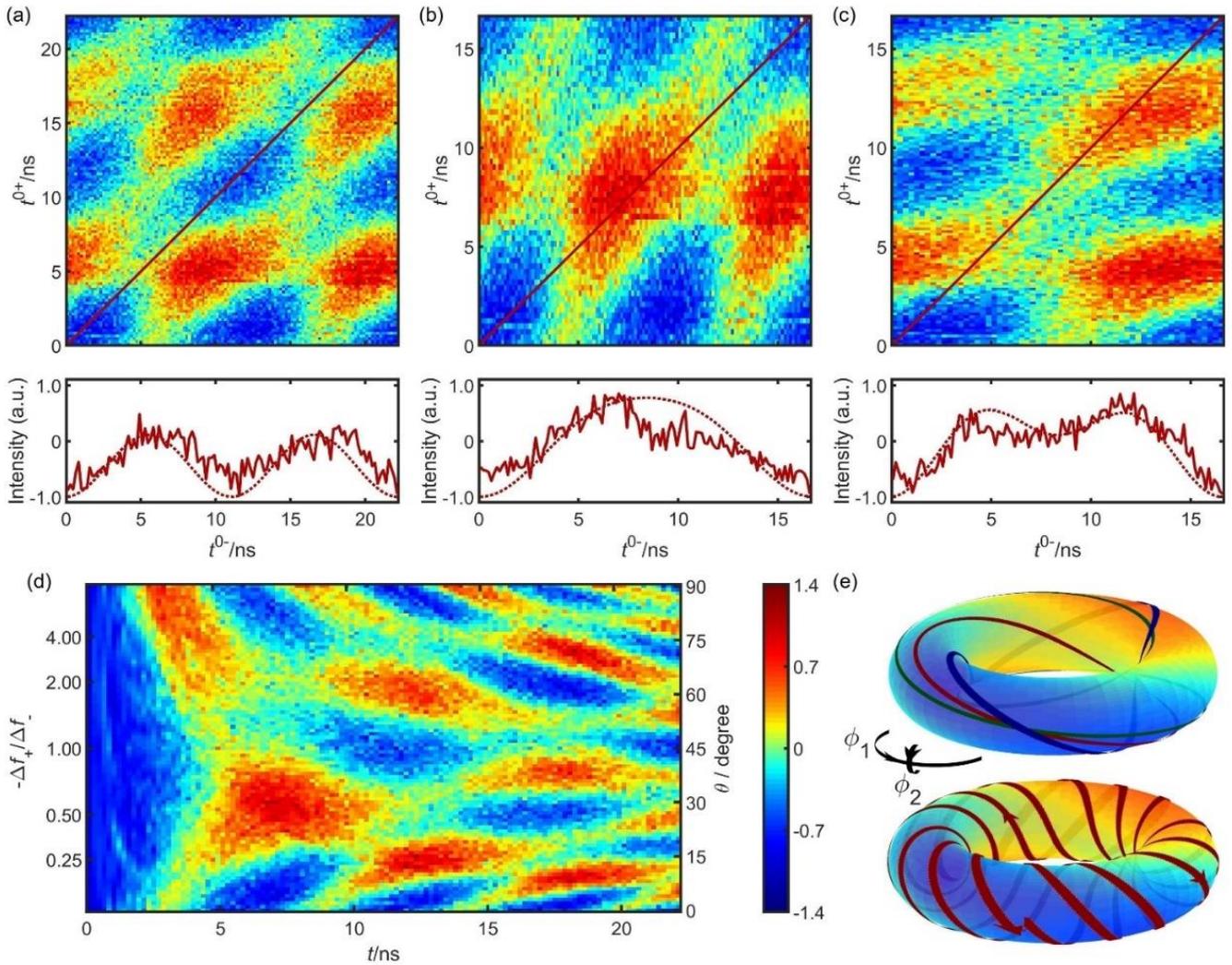

**Figure 4.** Interpretation of the quantum phase interference pattern as a map of evolution. (a-c) The interference pattern of $|\psi_2\rangle$ reinterpreted in time coordinates $t^{0\pm} = \pm\phi_\pm/\Delta f_\pm$ with $\Delta f_+/\Delta f_-$ = (a) 1, (b) 0.5 and (c) 2. The red line in each case marks the path of evolution through time in each condition, along which the change of $M^{0+}$ is shown in the lower panels, with solid and dashed lines giving experimental and simulated results. (d) A summation of all evolution paths with different $\Delta f_+/\Delta f_-$, each represented by a horizontal section of this graph. The vertical axis is spanned evenly by $\theta = \arctan(\Delta f_+/\Delta f_-)$. (e) Paths of evolution represented by curves on a torus, with blue, red and green curves for $\Delta f_+/\Delta f_-$ = 0.5, 1 and 2 in the upper part, and the incommensurate case for irrational $\Delta f_+/\Delta f_-$ in the lower part. The tori are colored according to simulated data for clarity.

## Discussion

Utilizing various advantages of the fullerene, we used the photoexcited triplet of $C_{70}$ to demonstrate the coherent manipulation of a 3-level quantum system. Superposition states $|\psi_1\rangle = (|+\rangle + |-\rangle)/\sqrt{2}$ and $|\psi_2\rangle = (|+\rangle + |0\rangle + |-\rangle)/\sqrt{3}$, characteristic of a qutrit, were prepared, characterized and observed for phase interference, which is further explained as a map of evolution. This shows our capability to fully control this molecule-based $S = 1$ electron spin state, in both magnitudes and phases of superposition, which is required in applications such as Grover's algorithm.[17] Its multilevel nature means not only the possibility of encoding more information within a molecular unit, but also the emergence of phenomena such as incommensurate evolution not present in qubits. To the best of our knowledge, this arbitrary coherent manipulation of an electron spin qutrit is for the first time realized on a molecular basis, which paves the way toward scalable and addressable assemblies of multilevel quantum systems. This reminds us that high spin magnetic molecules such as $Mn^{2+}$ and $Gd^{3+}$ complexes might also be worth more attention, for their multilevel nature is promising for application once desirable dynamic property can be achieved.

## Methods

**Sample preparation.** The sample was prepared by dissolving high purity $C_{70}$ (from 1-Material) in $d^8$-toluene (from J&K, 99.5% D) to form a solution about 0.1 mg/ml. The solution was then degased, sealed in argon atmosphere and cooled rapidly to 5 K by the liquid helium cryostat to form a glassy-state solid solution.
**DFT calculation.** The spin density was calculated with B3LYP/6-31g* by Gaussian 09w.

**Quamtum state manipulation.** The preparation of the superposition states, tomography of their density matrices and the quantum phase interference experiments were conducted in the pulse EPR setup. The off-resonant microwave pulses was generated with a arbitrary wave generator (AWG).

## Data availability

The data are available from the authors upon reasonable request.

**Keywords:** fullerene • magnetic properties • qutrit • quantum phase interference • quantum computation

## Acknowledgements

This research is supported by National Basic Research Program of China (2017YFA0204903 and 2018YFA0306003), National Natural Science Foundation of China (21822301, 51802346), and Beijing Academy of Quantum Information Sciences (Y18G23)

## Author Contributions

Y.X.W. and Z.L. proceeded the experiments, Y.X.W., in assistance with S.Z., prepared the manuscript. Y.H.F. carried out the DFT calculation. S.D.J. designed the experiments. S.D.J. and S.G. conceived the project. All authors revised the manuscript. Y.X.W. and Z.L. contributed equally to this work.

## Competing Interests

The authors declare that there are no competing interests.

## Table of Contents graphical abstract

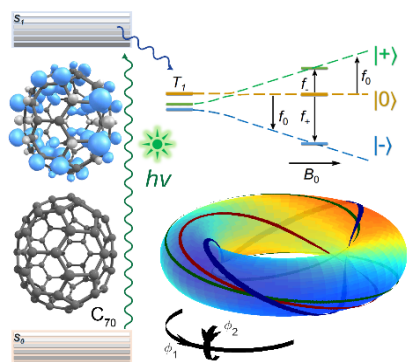

The photoexcited triplet of a fullerene is demonstrated as a molecular qutrit. The three-sublevel system can be prepared into superposition states with arbitrary amplitudes and phases, and exhibited intrinsic quantum phase interference behaviour.